\documentclass[10pt, letterpaper]{article}
\usepackage{amsfonts,amsmath,amssymb,mathrsfs,verbatim}

\font\tenmsy=msbm10
\font\sevenmsy=msbm10 at 7pt
\font\fivemsy=msbm10 at 5pt
\newfam\msyfam % family 11
\textfont\msyfam=\tenmsy
\scriptfont\msyfam=\sevenmsy
\scriptscriptfont\msyfam=\fivemsy
\def\blackB{\fam\msyfam\tenmsy}

\def\M{{\mathcal M}}
\def\Z{{\mathcal Z}}
\def\W{{\mathcal W}}

\def\NN{{\blackB N}}

\let\R\rangle
\let\l\left
\let\r\right

\let\da\dagger

\def\text#1{\quad\hbox{#1}\quad}

\def\la{\lambda}
\def\e{\epsilon}

\def\muh{{\hat \mu}}

\def\A{{\cal{A}}}

\def\rh{{\hat \rho}}
\def\lah{{\hat \lambda}}

\def\om{{\omega}}
\def\y{{\infty}}

\def\rw{\rightarrow}

\def\R{\rangle}

\def\rw{{\rightarrow}}

\begin{document}

\vskip-6cm

%vs $\W_k^{(k+1,k+3)}$ 
\title{\vskip0pt {\bf The $\W_k$ structure of the $\Z_k^{(3/2)}$ models}}
%\vskip18pt
%\title{\vskip0pt {\bf  $\Z_k^{(3/2)}\simeq \W_k(k+1,k+3)$}}

\smallskip
\author{ \bf{ 
P. Mathieu}\ 
\bigskip\\
D\'epartement de physique, de g\'enie physique et d'optique,\\
Universit\'e Laval,
Qu\'ebec, Canada, G1K 7P4,\\
pmathieu@phy.ulaval.ca}

\maketitle

% \begin{abstract}

%\vskip18pt
\centerline{{\bf ABSTRACT}}
%\vskip18pt

Generalized $\Z_k^{(r/2)}$ parafermionic theories -- characterized by the dimension $(r/2)(1-1/k)$ of the basic parafermionic field -- provide potentially interesting quantum-Hall trial wavefunctions. Such wavefunctions 
reveal 
a $\W_k$ structure. This suggests the equivalence of (a subclass of) the $\Z_k^{(r/2)}$ models and the $\W_k(k+1,k+r)$ ones. 
This  is demonstrated here for $r=3$ (the gaffnian series).
%the simplest generalized parafermionic models, the $\Z_k^{(3/2)}$ models
The agreement of the parafermionic and the $\W$ spectra relies on the prior determination of the field identifications in the parafermionic case.

\section{Introduction}
%==============================================================================

An interesting application of conformal field theory to the fractional quantum Hall effect is that CFT correlators can be used as trial wavefunctions  \cite{MR}.
In this way, the Read-Rezayi states \cite{RR} are related to the usual $\Z_k$ parafermionic theories \cite{ZFa}. A natural extension is to consider the $\Z_k^{(r/2)}$ generalized parafermionic models to generate  new classes of trial wavefunctions. Here the parameter $r$ is an  integer $\geq 2$ that specifies 
the conformal dimension of the parafermionic fields
$\psi_n$ to be  \cite{ZFa}
\begin{equation} \label{modelb}
h^{(r/2)}_{\psi_n} = {r n(k-n)\over 2k}\,\qquad \text{(where $r k\in 2\NN$),}
\end{equation}
with $\Z^{(1)}_k\equiv \Z_k$.\footnote{The $\Z_k^{(2)}$ models have been studied in \cite{ZFc} for $k=3$ and in \cite{Dot} for arbitrary $k$ (but in both cases,  only for the unitary series), while the $\Z_3^{(4)}$ model has been constructed  in \cite{DotS}.}
The $\Z_k^{(3/2)}$ parafermionic theories have been  introduced in \cite{JM32}; their $k=2$  correlators are related to the so-called gaffnian states \cite{SRC}.\footnote{As shown below, the  $\Z_2^{(3/2)}$ model is indeed equivalent to the $\M(3,5)$ minimal model, the CFT underlying the gaffnian states. The later is also the first member of yet another parafermionic sequence, the so-called graded parafermionic theories, related to the coset $\widehat{osp}(1,2)_k/\widehat{u}(1)$ \cite{CRS}.}

Up to exponential terms, the (bosonic) quantum-Hall wavefunction  is a symmetric polynomial. Wavefunctions 
with prescribed parafermionic-type clustering properties (vanishing with power $r$ when $k+1$ coordinates approach each other) happen to be given by particular Jack polynomials  with parameter $\alpha=-(k+1)/(r-1)$ and partitions with difference $r$ at distance $k$ \cite{BH,BGS}. These are precisely the Jack polynomials that were conjectured to be related to the minimal models $\W_k(k+1,k+r)$ \cite{Fetal}. Evidences for this Jack-$\W$ relationship are presented  in \cite{BHa}. This connection has been further substantiated in \cite{BGS}, where in particular the central charge of the CFT related to these special Jack wavefunctions is computed directly. As expected, the underlying CFT displays the clear characteristics of a $\Z_k^{(r/2)}$ theory.

These considerations suggest the 
general CFT equivalence 
\begin{equation}\label{zwgen}
\Z_k^{(r/2)}\simeq \W_k{(k+1,k+r)} ,\end{equation}
(already hinted at in \cite{JM32}). This relation incorporates the well-established unitary $\W$ representation of the usual $\Z_k$ models \cite{Nar}.
This correspondence is also verified for $k=2$ \cite{JM3p} where (\ref{zwgen}) reduces to
$\Z_2^{(r/2)}\simeq \M(3,r+2)$,
in which case, the parafermion $\psi_1$, of dimension $r/4$, is identified with the minimal-model primary field $\phi_{2,1}$.
However, for $r\geq 4$ and $k>2$, (\ref{zwgen}) is  incomplete since the central charge of the parafermionic theory does not appear to be fixed. This relation should thus be understood as restricted to a particular non-unitary sector (or minimal series) of the parafermionic models, a sector selected by the assumed properties of
the basic parafermionic correlation function. Alternatively, it states 
that these particular $\W_k$ models have a reformulation in terms of  $\Z_k^{(r/2)}$ parafermions.

Apart from the $k=2$ or $r=2$ cases, there is an another instance where the equivalence (\ref{zwgen}) should be satisfied without restriction
and this is when $r=3$.
As for their $r=2$ counterparts, the central charge of the $\Z_k^{(3/2)}$ models  is uniquely fixed by associativity; it reads
\begin{equation}\label{centre}
c=-{3(k-1)^2\over (k+3)}\; . \end{equation}
This is the first indication of the equivalence:
\begin{equation}\label{zwiden}
\Z_k^{(3/2)}\simeq \W_k{(k+1,k+3)} \;.\end{equation}
This relation was pointed out in \cite{JM32} and claimed to hold in the simplest cases, albeit without the presentation of a detailed supporting analysis.
This statement relied on the equality of the central charges and the fact that the $\W$ primary fields 
were contained in the set of parafermionic primary fields and the top fields of different charge modulo $k$ among the parafermionic descendants.
However, this verification did not rest on the analysis of the field identifications in the parafermionic models. This is remedied here, where at first the field identifications in the $\Z_k^{(3/2)}$ model are obtained, and then used to verify the perfect correspondence between the spectra of the two theories in (\ref{zwiden}).

\section{The $\Z_k^{(3/2)}$ parafermionic theory}

\subsection{Structure of the $\Z_k^{(3/2)}$  algebra}

Let us first briefly review the basic elements of the  $\Z_k^{(3/2)}$  parafermionic models \cite{JM32}.
As for any $\Z_k$ theory, there are $k$ sectors labeled by
integers $t=0,\cdots , k-1$. 
The mode decomposition of $\psi_1$ in the sector $t$, reads
\begin{equation}
 \psi_1(z)  = \sum_{m=-\y}^\y
z^{-\frac{t}{k}-m-1}A_{m+1+\frac{t}{k}-\frac32+\frac{3}{2k}}=\sum_{m=-\y}^\y
z^{-\la q-m-1}A_{m-\frac12+\la(1+q)} \; , \label{modep}\end{equation}
where
\begin{equation}
t={3 q\over 2}\; , \qquad \la= {3\over 2k}\; . \end{equation}
A similar
expression holds for the decomposition of
  $\psi^\dagger_1$ with $q\rw-q$.
 The charge $q$ is normalized by setting that $A$ to be 2 (or its $t$ value is 3).
  It is convenient to write
\begin{equation}
A_{u+\la(1+q)}= \A_u\quad \text{and}\quad \A^\da_{u+\la(1-q)}=\A^\da_u , \end{equation}
where $u$ is half-integer.
With this notation the commutation
relations reads (when acting  on a state in the $t$ sector):
\begin{equation} \label{comisaa}
  \sum_{l=0}^{\infty} 
  \binom{l+2\la-1}{l}
  \left[ \A_{n-l-\frac12}
\A^\dagger_{m+l+\frac12} - \A_{m-l+\frac12}^\dagger  \A_{n+l-\frac12}
  \right]
  = \left[  -{(k+3)\over k(k-1) }  L_{n+m} + {1 \over 2}
(n+{\la q})(n-1+\la q) \delta_{n+m,0} \right].
\end{equation}
(The coefficient of the Virasoro mode is actually $2h_1/c$, where  $h_1=3(k-1)/(2k)$, the dimension of $\psi_1$; associativity fixes $c$ to the value (\ref{centre}), which has been substituted here.)
In addition, one has
\begin{equation}
  \sum_{l=0}^{\infty} 
    \binom{l-2\la-1}{l}
    %C^{(l)}_{2\la}
     \left[ \A_{n-l-\frac12}
\A_{m+l+\frac12} - \A_{m-l+\frac12}  \A_{n+l-\frac12}
  \right]  = 0 \;,
\label{comisaTwo}\end{equation} 
with an identical expression
with $\A$ replaced by $\A^\dagger$.

\subsection{Aspects of the representation theory}

The highest-weight states  are  defined by the conditions:
\begin{equation}
\A_{m+\frac12}|{\rm hws}\R  = \A_{m+\frac12}^\dagger|{\rm hws}\R= 0 \quad
{\rm for}\,\;\; m\geq 0\;. \label{highestW}\end{equation}
They  are characterized by $t$ and a further quantum number $s$ (called $r$ in \cite{JM32}):
\begin{equation}|{\rm hws}
\rangle \equiv |t,s\rangle \;. \end{equation}
The parameter
$s$ enters in the expression of the basic singular vectors
\begin{equation}\label{singA}
\text{(i):} (\A_{-\frac12})^{s+1}|{t,s}\R \qquad\text{and}\qquad \text{(ii):}( \A^{\dagger}_{-\frac12} )^{t+s+1} |{t,s} \rangle ,
\end{equation}
which obey the highest-weight conditions (\ref{highestW}).  Quite remarkably, these conditions
fix the conformal dimensions  of the highest-weight states to be
\begin{equation}
h_{t,s} = -{ k (k-2s-t-1)(2s+t) +t^2 \over 2k(k+3)}\;.
\label{confDim}\end{equation}
The corresponding primary field will be denoted $\phi_{t,s}$. The state with $t=s=0$ has $h=0$, so that it can be identified with the vacuum (i.e., $\phi_{0,0}=I$). In addition, we find the singular vectors
\begin{align}
\text{(i):} ( \A_{-\frac32} )^{k-t-2s+1} ( \A_{-\frac12} )^{s} |t,s \rangle
\qquad \text{and}\qquad \text{(ii):}( \A^{\dagger}_{-\frac32} )^{k-t+1}( \A^{\dagger}_{-\frac12} )^{t+s} |t,s \rangle ,
\label{singB}\end{align}
as solutions of the weaker conditions
\begin{equation}
\A_{\frac32} |\chi \rangle= 0 = \A^{\dagger}_{\frac32} |\chi \rangle ,
\end{equation}
instead of (\ref{highestW}) (which signals that these singular vectors are actually descendants of a lower dimensional singular vector involving Virasoro modes \cite{JM32}). 
 These conditions will be sufficient for our purpose.

By definition, $s$ has to be a non-negative integer
 while from (\ref{singB}), we deduce that $0 \leq s \leq
(k-t)/2$.  However, the bounds:
\begin{equation}\label{trb}
0\leq t\leq k-1\, \quad \text{and}\quad 0\leq s\leq Ê\l\lfloor\frac{k-t}2\r\rfloor\, 
\end{equation} 
(where $\lfloor x\rfloor$ is the integer part of $x$),
are not optimal: not each allowed value of $(t,s)$ corresponds to an independent field.

\subsection{Field identifications}

In order to obtain a set of field identifications, we use as a guiding principle the observation that for the usual $\Z_k$ models, these  involve fields associated to states at the inner border of the sequences defining the singular vectors \cite{JM}. In this way, we readily obtain the identifications:
\begin{subequations}
\begin{align}
(\A_{-\frac32})^{k-t}|t,0\R &\sim |k-2t,t\R\label{FIa}\\
(\A_{-\frac32})^{k-2s}   (\A_{-\frac12})^{s}|0,s\R &\sim |k-3s,s\R \label{FIb}\\
(\A^\da_{-\frac12})^{t+s}|t,s\R &\sim |k-2t-3s,s+t\R\label{FIc}
\end{align}
 \end{subequations}   
   with the requirements that the two entries $t',s'$ specifying a state need to be positive integers (e.g, in the second case, we require $k-3s\geq 0$). In each case, the equality of the dimensions and the charges (modulo $k$) is verified. 

The lhs of the two first identifications are two special cases of the one-before-last state of the sequence (\ref{singB}i) (for $s=0$ and $t=0$ respectively). Similarly, the lhs of (\ref{FIc}) is related to the penultimate state in the sequence (\ref{singA}ii). The remaining two sequences, (\ref{singA}i) and (\ref{singB}ii), do not lead to state identifications.

The identifications (\ref{FIa})-(\ref{FIc}) are established as follows: one acts on the lhs by either $\A_{-\frac32}$ (in the first two cases) or $\A^\da_{-\frac12}$ (in the third one) to generate a singular vector that is set equal to zero. The same action on the rhs becomes either $\A_{\frac12}$ or $\A^\da_{\frac12}$ respectively, which is nothing but a highest-weight condition. For instance, in the second case, one has
\begin{equation}
 \A_{-\frac32}(\A_{-\frac32})^{k-2s}   (\A_{-\frac12})^{s} |0,s\R =A_{-\frac32+\frac3{2k}(1+\frac23(3k-3s))}(\A_{-\frac32})^{k-2s}   (\A_{-\frac12})^{s}|0,s\R =0.
\end{equation}
When acting on the rhs, this becomes
\begin{equation}
   A_{-\frac32+\frac3{2k}(1+\frac23(3k-3s)) }|k-3s,s\R=   A_{-\frac32+2+\frac3{2k}(1+\frac23(k-3s))} |k-3s,s\R= \A_{\frac12}  |k-3s,s\R=0.
\end{equation}

In addition to these series of identifications, one has the following sequence
\begin{equation}\label{zkinv}
(\A_{-\frac32})^{k-t-2s} (\A_{-\frac12})^{s}|t,s\R \sim(\A^\da_{-\frac12})^{t+s}|t,s\R,
\end{equation}
which merely reflects the $\Z_k$ cyclic symmetry, expressed here as a state identification  in its most general form. In particular, it implies that the identification (\ref{FIa}) is a special case of (\ref{FIc}).
 The condition (\ref{zkinv}) also entails $(\A_{-\frac32})^{k}|0,0\R \sim|0,0\R$ and 
$(\A^\da_{-\frac32})^{k-t}(\A^\da_{-\frac12})^t|t,0\R \sim|t,0\R$.

\section{Generalities concerning the $\W_k{(p',p)}$ model}

In view of establishing the equivalence (\ref{zwiden}), let us first recall some results on
$\W_k{(p',p)}$ models \cite{FL}. The primary fields
$\phi_{\{\lah,\muh\}}
$are labeled by two integrable $\widehat {su}(k)$  weights  $\lah$
and $\muh$ at respective level
$p'-k$ and
$p-k$ (with say $p>p'$).
Their conformal dimension reads
\begin{equation}
h_{\{\lah,\muh\}}=
{|p(\la+\rho)-p'(\mu+\rho)|^2-(p-p')^2|\rho|^2\over 2 pp'}\;, \end{equation}
where $\la$ stands for the finite weight associated to $\lah$ and
$\rho$ is the Weyl vector (we follow the notation of \cite{CFT}):
\begin{equation}\lah = \sum_{i=0}^{k-1}\la_i {\widehat\omega}_i =  [\la_0,\la_1,\cdots,\la_{k-1}], \qquad
\rh = \sum_{i=0}^{k-1}{\widehat\omega}_i =  [1,1,\cdots,1],\end{equation} 
Recall also the $\W$ field identifications \cite{FL}:
\begin{equation}\label{wfi}
\{\lah,\muh\} \sim \{a\lah,a\muh\}\sim \cdots \sim
\{a^{k-1}\lah,a^{k-1}\muh\}\;,
\end{equation}
where $a$ is the basic $\widehat {su}(k)$ automorphism that permutes
the Dynkin labels:
\begin{equation}
a\, [\la_0,\la_1,\cdots,\la_{k-1}]=[\la_{k-1},\la_0,\cdots,\la_{k-2}].
\end{equation}

The central charge of the $\W_k{(p',p)}$ models is
\begin{equation}
c= (k-1)\left(1-{k(k+1)(p-p')^2\over pp'}\right)
\end{equation}
and with $(p',p)= (k+1,k+3)$, this reduces to (\ref{centre}). This completes the first step of the verification of (\ref{zwiden}).

In preparation for the next step, note that for $(p',p)= (k+1,k+3)$, $\lah\in P_+^{1}$ and $\muh\in P_+^{3}$, with $P_+^{\,m}$ denoting the set of integrable weights at level $m$. Thanks to the field identifications (\ref{wfi}), we can choose the  field representatives to be all
of the form: 
\begin{equation}
{\{[1,0,\cdots,0],\muh\}}= \{{\widehat\omega}_0,\muh\} ,
\end{equation}
with
$\muh$ running over the complete set $P_+^3$. We will thus designate the $\W$ fields
solely by $\muh$. Note that by acting with $a^n$ only on $\muh$ does produce distinct fields for $n<k$

\section{$\Z_k^{(3/2)}$  vs the  $\W_k{(k+1,k+3)}$ model}

The next step amounts to compare the spectrum of the two theories. Since it is not clear at once how the $\W$ primary fields get reorganized in terms of parafermionic families, it is more appropriate at this point to first perform an explicit analysis  for  the lowest values of $k$, e.g., $k=2,4$ and 6.

\subsection{$k=2$}
The $\W_2{(3,5)}$ model is nothing but the minimal model $\M(3,5)$ with central charge $-3/5$. With  their conformal
dimension indicated by an attached subscript, the $\W_2{(3,5)}$ fields,
designated by $\muh$, are:
\begin{equation}
[3,0]_0\,,\qquad [0,3]_{\frac34}\,,\qquad [2,1]_{\frac{-1}{20}}\,,\qquad [1,2]_{\frac15}.
\end{equation}
States have been ordered in orbits of $a$ (here of length 2).
These dimensions match those of the parafermionic states:
\begin{equation}\label{k2p}
|0,0\R\, ,\qquad\A_{-\frac32}|0,0\R \,,\qquad |1,0\R\,, \qquad\A_{-\frac32}|1,0\R. 
\end{equation}
 We see that the parafermion acts as the simple current (as $a$).
To be fully explicit: the state $|0,0\R$ has already been identified with the vacuum, while  for the state $|1,0\R$, we have
\begin{equation}
h_{t,0}= -{t(k-t)(k-1)\over 2k(k+3)}\quad\xrightarrow{(k=2,\,t=1)}\quad  h_{1,0}= -\frac1{20}.
\end{equation}
The dimensions of the second and the fourth states are  computed as
\begin{equation}
\A_{-\frac32}|0,0\R = A_{-\frac32+\frac43}|0,0\R=A_{-\frac34}|0,0\R\quad \text{and}\qquad \A_{-\frac32}|1,0\R = A_{-\frac32+\frac43(1+\frac23)}|1,0\R=A_{-\frac14}|1,0\R,
\end{equation}
giving then respectively $3/4$ and  $1/4+h_{1,0}=1/5$.

These correspondences suggest that the first and the third states in (\ref{k2p})
are the only two parafermionic primary fields. However, the bounds (\ref{trb}) allow also the solution: $t=0,s=1$.
From the expression of $h_{t,s}$ in (\ref{confDim}),  we have
\begin{equation}
h_{0,s} = -{s (k-2s-1) \over (k+3)}\quad\xrightarrow{(k=2,\, s=1)}\quad
h_{0,1}={1\over 5}. \label{confDimor}\end{equation} 
This state is identified as follows (cf. (\ref{FIa}) with $k=2,\,t=1$):
$|0,1\R\sim \A_{-\frac32}|1,0\R$.
Its $\A_{-\frac12}$ `descendant' is
\begin{equation}
\A_{-\frac12}|0,1\R = A_{-\frac12+\frac43(1+0)}|0,1\R=A_{\frac14}|0,1\R \; : \quad h=-\frac14+\frac15=-\frac1{20},
\end{equation}
so that $\A_{-\frac12}|0,1\R\sim |1,0\R.$
This exhausts the spectrum.

\subsection{$k=4$}

The different primary fields of the $\W_4{(5,7)}$ model are:
\begin{align}
&  [3,0,0,0]_0 &\quad & [0,3,0,0]_{\frac98} &\quad & [0,0,3,0]_{\frac32} &\quad
&[0,0,0,3]_{\frac98 }\cr
& [2,1,0,0]_{\frac{-9}{56}}&  &[0,2,1,0]_{\frac57}& & [0,0,2,1]_{\frac{47}{56}}& &[1,0,0,2]_{\frac3{14}}\cr
& [2,0,1,0]_{\frac{-3}{14}}&  &[0,2,0,1]_{\frac{23}{56}}& & [1,0,2,0]_{\frac27}&   &[0,1,0,2]_{\frac{23}{56}}\cr
& [2,0,0,1]_{\frac{-9}{56}} & & [1,2,0,0]_{\frac 3{14}} &  & [0,1,2,0]_{\frac{47}{56}}&   &[0,0,1,2]_{\frac57}\cr
 &[1,1,0,1]_{\frac{-1}7} & & [1,1,1,0]_{\frac{-1}{56}}&   & [0,1,1,1]_{\frac5{14}}&  &[1,0,1,1]_{\frac{-1}{56}}
\end{align}
The $\W$ fields have been organized (horizontally) in orbits of the outer automorphism with $a^n$ ($n=0,1,2,3$).
% acting solely on its  $\muh$-th component.  
For each orbit, the lowest-dimensional field has been placed  at the left-most position. Again, this action of the outer automorphism is that of a simple current which is the parafermion itself. 

The states in the first row are thus described by the string
\begin{equation}
(\A_{-\frac32})^n|0,0\R\qquad {\rm with} \quad |0,0\R \sim [3,0,0,0] \quad {\rm and} \quad n=0,1,2,3\;.\end{equation}
We can thus associate the four states on the top row with the fields $\psi_n$, with $n=0,1,2,3$ respectively.
Let us now compare the conformal dimensions of the parafermionic highest-weight  state with that of the left-most state in the following three rows under the assumption that these all have $s=0$:
\begin{equation}
h_{t,0}= -{t(k-t)(k-1)\over 2k(k+3)}\quad\xrightarrow{(k=4)}\quad  h_{1,0}=h_{3,0}= -{9\over 56}\;, \quad h_{2,0}=-{3\over 14}\;. \end{equation}
That shows that we can identify the spin field $\phi_{t,0}$ with  the $\W$ field whose finite weight $\mu$ is the fundamental weight $\omega_t$. The other states in these rows are described by the orbits: 
\begin{equation}
(\A_{-\frac32})^n|t,0\R \quad {\rm with}\quad 0\leq n\leq
4-t\qquad {\rm and}\quad (\A_{-\frac12}^\dagger)^m|t,0\R\quad {\rm with}\quad 1\leq m\leq t-1,
\end{equation}
 which are within the bounds fixed by the singular vectors (\ref{singA}) and (\ref{singB}).
When a singular vector is hit with the first sequence (namely when the Dynkin label 1 precedes the 2), one restarts with the other one. The first descendant state generated from  the second string is  the one at the far right state. Indeed $\A^\da$ acts as $a^{-1}$.
Note that the above bounds on $n$ and $m$ could be modified as $0\leq n\leq 3-t$ and $1\leq m\leq t$ due to the field identification (\ref{zkinv}).

All the states considered so far have $s=0$ and there is no room left  for further such states.  But we still have to properly identify the states in the last row. Let us then see how we could interpret its left-most state, of dimension $-1/7$, as a highest-weight states $|t,s\R$ with $s\not=0$. The first step is to fix the value of $t$ using its additive conservation (modulo $k$) in fusions. Since
\begin{equation}[2,1,0,0]\times [2,0,0,1]\supset [1,1,0,1] ,
\end{equation}
(meaning that $[1,1,0,1]$ appears in the fusion rule $\phi_{1,0}\times \phi_{3,0}$),
% and that the sector label  $t$ is additive modulo $k=4$, 
we conclude that  $[1,1,0,1]$ has $t=0$.
Enforcing $h_{0,s}=-1/7$ fixes $s=1$.
 We thus
identify the state $[1,1,0,1]$ with $|0,1\R$.  
Its orbit is checked to be correctly described by  the sequence 
\begin{equation}
(\A_{-3/2})^n  \A_{-1/2}|0,1\R \;\quad\text{with $n=0,1,2$ respectively.} \end{equation}

In this $k=4$ example, we see by inspection that the value of $t$ of a parafermionic primary field is nicely related to the finite Dynkin labels of the corresponding $\W$ field $\muh$  as
$t = \sum_{i=1}^{3}i \mu_i$ mod 4. For parafermionic descendant, the value of $t$ is augmented  by 3 times the number of $\A$ modes or $-3$ times the number of $\A^\da$ modes, again with the addition understood modulo $k$.
This  is again in agreement with the above formula for the corresponding $\muh$ field. For instance, with $(\A_{-\frac32})^3|1,0\R\sim [1,0,0,2]$, $ t= 1+9=2$ (mod 4) in the parafermionic description and as a $\W$ field, it is $t=6 =2$ (mod 4).
The expression $\sum_{i=1}^{3}i \mu_i$ (mod 4)
is the natural Lie algebraic interpretation of the parafermionic charge since  it defines to the so-called $su(4)$ congruence classes \cite{LPa} (see e.g. \cite{CFT} chap. 13), which are additively conserved in tensor products, hence in fusion rules. The generalization to all $k$ is obviously
\begin{equation}\label{tkg}
t = \sum_{i=1}^{k-1}i \mu_i \quad {\rm mod}\;  k.
\end{equation}

Back to the spectrum analysis of our $k=4$ example. Not all values of $|t,s\R$ have been related to $\W$ primary fields. However, since there are no more $\W$ primary fields, the remaining $|t,s\R$ states, namely, $|0,2\R,\,|1,1\R$ and $|2,1\R$, should be related to states already obtained. Indeed, one has
\begin{align}
&|0,2\R\sim(\A_{-\frac32})^2|2,0\R\qquad \qquad\text{[by (\ref{FIa})]} \nonumber\\
&|1,1\R\sim(\A_{-\frac32})^2\A_{-\frac12}|0,1\R\qquad  \text{[by (\ref{FIb})]}\nonumber\\
&|2,1\R\sim(\A_{-\frac32})^3|1,0\R\qquad \qquad \text{[by (\ref{FIa})].}
\end{align}
We thus find a perfect agreement between the spectra of the  $\Z_4^{(3/2)}$ and the  $\W_5{(5,7)}$ models.

\subsection{$k=6$}

The set of $\W_6(7,10)$ fields can be organized in terms of 10 orbits.
 The fields  specified by the  lowest dimensional member of its orbit (except in the penultimate case) -- with the comas between the Dynkin labels omitted --, followed by the dimensions of 6 orbit members are:
\begin{align}
&[300000]:&&0&& \tfrac54 && 2&& \tfrac94 &&2 &&\tfrac54&& (\A_{-\frac32})^5|0,0\R\nonumber\\
&[210000]:&&\tfrac{-25}{108} && \tfrac{23}{27} && \tfrac{155}{108}&& \tfrac{41}{27} &&\tfrac{119}{108} &&\tfrac5{27} && (\A_{-\frac32})^5|1,0\R\nonumber\\
&[201000]:&&\tfrac{-10}{27} && \tfrac{59}{108} && \tfrac{26}{27}&& \tfrac{95}{108} &&\tfrac{8}{27} &&\tfrac{23}{108} && 
(\A_{-\frac32})^4|2,0\R\,,\; \A^\da_{-\frac12}|2,0\R\nonumber\\
&[200100]:&&\tfrac{-5}{12} && \tfrac{1}{3} && \tfrac{7}{12}&& \tfrac{1}{3} &&\tfrac{7}{12} &&\tfrac{1}{3} && 
(\A_{-\frac32})^3|3,0\R\, ,\; (\A^\da_{-\frac12})^2|3,0\R\nonumber\\
&[200010]:&&\tfrac{-10}{27} && \tfrac{23}{108} && \tfrac{8}{27}&& \tfrac{95}{108} &&\tfrac{26}{27} &&\tfrac{59}{108} && 
(\A_{-\frac32})^2|4,0\R\, ,\; (\A^\da_{-\frac12})^3|4,0\R\nonumber\\
&[200001]:&&\tfrac{-25}{108} && \tfrac{5}{27} && \tfrac{119}{108}&& \tfrac{41}{27} &&\tfrac{155}{108} &&\tfrac{23}{27} && 
\A_{-\frac32}|5,0\R\, ,\; (\A^\da_{-\frac12})^4|5,0\R\nonumber\\
&[110001]:&&\tfrac{-1}{3} && \tfrac{-1}{12} && \tfrac{2}{3}&& \tfrac{11}{12} &&\tfrac{2}{3} &&\tfrac{-1}{12} && 
(\A_{-\frac32})^4\A_{-\frac12}|0,1\R\nonumber\\
&[101001]:&&\tfrac{-37}{108} && \tfrac{-7}{27} && \tfrac{35}{108}&& \tfrac{11}{27} &&\tfrac{-1}{108} &&\tfrac{2}{27} && 
(\A_{-\frac32})^3\A_{-\frac12}|1,1\R\, ,\; \A^\da_{-\frac12}|1,1\R\nonumber\\
&[100101]:&&\tfrac{-7}{27} && \tfrac{-37}{108} &&\tfrac{2}{27}&& \tfrac{-1}{108}&& \tfrac{11}{27} &&\tfrac{35}{108} &&
(\A_{-\frac32})^2\A_{-\frac12}|2,1\R\, ,\; (\A^\da_{-\frac12})^2|2,1\R\nonumber\\
&[101010]:&&\tfrac{-2}{9} && \tfrac{1}{36} &&  &&    &&  &&\ && 
\A_{-\frac12}|0,2\R\,
\end{align}
At the right of each row, we have given the corresponding parafermionic states as a sequence of operators acting on the parafermionic primary state.
When two sequences are written, we recall that $\A^\da$ acts as $a^{-1}$ so that the outer automorphism action is toward the right. For instance we have:
\begin{align}
&(\A_{-\frac32})^2|2,0\R= A_{-\frac5{12}}A_{-\frac{11}{12}}|2,0\R\sim [002010],&&h=\frac5{12}+\frac{11}{12}-\frac{10}{27}=\frac{26}{27},&& t= 2+6=2 \, ({\rm mod}\,6)&\nonumber\\
&(\A^\da_{-\frac12})^2|2,1\R= A^\da_{-\frac1{12}}A^\da_{-\frac7{12}}|2,1\R\sim [010110],&&h=\frac1{12}+\frac7{12}-\frac{7}{27}=\frac{11}{27},&&t=2-6=2\,({\rm mod}\,6).&
\end{align}
The $t$ assignments are in agreement with  the expression (\ref{tkg}).
The last row displays a feature encountered for all values of $k$ that are multiple of 3: a short orbit that results from the existence of fixed points under the action  of some power of $a$ (here $a^2$).

The missing parafermionic primary states are taken  into account by the following identifications:
\begin{align}
&|4,1\R\sim(\A_{-\frac32})^5|1,0\R\qquad \qquad\text{[by (\ref{FIa})]} \nonumber\\
&|2,2\R\sim(\A_{-\frac32})^4|2,0\R\qquad \qquad\text{[by (\ref{FIa})]} \nonumber\\
&|0,3\R\sim(\A_{-\frac32})^5|3,0\R\qquad \qquad\text{[by (\ref{FIa})]} \nonumber\\
&|3,1\R\sim(\A_{-\frac32})^4\A_{-\frac12}|0,1\R\qquad  \text{[by (\ref{FIb})]}\nonumber\\
&|1,2\R\sim(\A^\da_{-\frac12})^2|1,1\R\qquad \qquad \text{[by (\ref{FIc})].}
\end{align}
This completes the verification of the spectrum equivalence of the $\Z_6^{(3/2)}$ and the $\W_6{(7,9)}$ models.

\subsection{Generic even $k$}

With $k$ generic, we can easily verify the following correspondences.
The parafermion $\psi_n$ is associated to the field whose finite weight
$\mu= 3\om_n$, whose conformal dimension is $(3n/2)(1-n/k)$. Similarly, the spin field $\phi_{t,0}$ corresponds to that with weight
$\mu= \om_t$, whose dimension  equals $h_{t,0}$. The description of its orbit in terms of parafermionic descendants is readily verified.

$\W_k$ primary fields with $0\leq \mu_i\leq 1$
($i=0,\cdots , k-1)$ (that is, with three affine Dynkin labels equal to
1)  correspond to parafermionic states with $s\not=0$. The corresponding parafermionic primary state has $\mu_0=1$; with
 $i,j$ standing for the positions of its non-zero finite Dynkin labels,
its value of $s$ reads
\begin{equation}
s= \frac12(k-t- |i-j|)\;.
\end{equation}
 This is easily checked to be integer: $k$ is even, $t$ is $i+j$ modulo $k$, and $|i-j|=i-j$ modulo 2.
This  expression for $s$ is verified in the previous case $(k=2,4,6)$. For $k=8$,  here are some correspondences:
\begin{align}
&[11000001]_{\frac{-5}{11}}: \; |0,1\R& \quad & [10100010]_{\frac{-6}{11}}: \; |0,2\R & \quad & [10010100]_{\frac{-3}{11}}: \; |0,3\R   \nonumber\\
& [10000101]_{\frac{-4}{11}}: \; |0,4\R &\quad & [10100001]_{\frac{-97}{176}}: \; |1,1\R & \quad & [10010001]_{\frac{-25}{44}}: \; |2,1\R \;.
\end{align}

Let us conclude by comparing the number of orbits in the $\W_k(k+1,k+3)$ model with the number of parafermionic primary fields (which should be equinumerous).
The number of orbits is given by (e.g., \cite{CFT} eq. (16.159)):
\begin{equation}\label{orb}
\l\lceil \frac{|P_+^3|}{k}\r\rceil = \l\lceil    \frac{(k+2)!}{k! \, 3!}\r\rceil = \l\lceil    \frac{(k+2)(k+1)}{6}\r\rceil,
\end{equation}
(where $\lceil x \rceil $ stands for the smallest integer larger than $x$). Inspection of the field identifications (\ref{FIb}) and (\ref{FIc}) (recalling that (\ref{FIa}) is a special case of (\ref{FIc})) indicates that the bounds on $s$ and $t$ that generate an independent set of fields are
\begin{equation}
0\leq s\leq \l\lfloor \frac{k}3\r\rfloor,\qquad 0\leq t\leq k-1-3s+\delta_{s,k/3}.
\end{equation}
Counting the number of allowed solutions to these inequalities (treating the three cases $k=3\ell+\e$, with $\e=0,1,2$, separately) gives precisely the number (\ref{orb}).

This analysis makes the equivalence (\ref{zwiden}) firmly established.

\noindent {\bf ACKNOWLEDGMENTS}

This work  is supported  by NSERC. I thank P. Jacob for his collaboration on \cite{JM32} and V. Gurarie for discussions concerning \cite{BGS}.

%%%%%%%%% section free basis%%%%%%%%%%%%
\end{document}